# Electron and Superconducting Properties of the *A*FeAs (*A*= Li, Na) Family Alkali-Metal Pnictides: Current Stage of the Research (mini-review)

T. E. Kuzmicheva[+ 1], S. A. Kuzmichev[*+]

[+] Lebedev Physcial Institute RAS, 119991 Moscow, Russia

[*] Faculty of Physics, Lomonosov Moscow State University, 119991 Moscow, Russia

The review is focused on one of the most exotic families of iron-based superconductors belonging to the *A*FeAs structural type, where *A* is alkali metal. We briefly concern physical and electron properties of the typical members of this family, LiFeAs and NaFeAs, discuss the theoretical models describing the multiple-gap superconducting state, and the experimental data available in literature. As well, we specify the main unsolved problems, that seem crucial for both the *A*FeAs family and for iron-based superconductors in general.

**1. Introduction.** Layered alkali-metal iron pnictides *A*FeAs have moderate critical temperatures $T_c$ up to 22 K and belong to a so called 111 structural family. Similarly to other iron-pnictide families, the crystal structure of the 111 family contains superconducting FeAs blocks separated by alkali metal blocks along the *c* - direction. The 111 family members are not so numerous: stable chemical structures are formed only with alkali metals having small atomic radius (Li and Na), whereas the possible substitutions are limited by a certain set of transition metals $Tm$ = Co, Ni, Cu, V, Rh, or alkali metal deficiency $A_{1-\delta}$FeAs. Nonetheless, the *A*FeAs compounds show extraordinary properties that are not typical for the majority of the iron-based high temperature superconductors (HTSC) and strongly depend on the chemical composition. Accordingly, *A*FeAs pnictides are of great fundamental interest.

Using "self-flux" technique, it is possible to grow high-quality *A*FeAs single crystals (as large as 1 cm in dimension) [1–3]. However, experimentalists usually have to overcome a number of troubles when probing the properties of these peculiar compounds. For example, when exposing LiFeAs crystal in open air, its critical temperature rapidly decreases, dropping to zero in about 10–20 min, whereas LiOH emerges between FeAs blocks. Since the 111 family crystals are naturally cleaved along the blocks of active alkali metal atoms, their surface degrades inevitably, in presence of even trace amounts of oxygen or water vapors. For a long time, nitrogen also reacts chemically with *A* FeAs. Although the bulk properties remain almost stable and insensitive to nitrogen, its presence is appears for the sample surface. Therefore, all preparations and the experiment have to be made in a "dry" vacuum or in a protective atmosphere.

Due to high quality of the cryogenic cleaves provided that the experiment is carried out correctly, availability of huge single crystals, and an absence of surface bands [4], the alkali-metal superconductors seem to be the best candidates for angle-resolved photoemission spectroscopy (ARPES) probes in order to determine the band structure features in a high resolution. Nonetheless, the above mentioned experimental troubles lead to the lack of experimental data on the properties of the 111 compounds measured by other techniques. In particular, the studies of the main characteristics of the superconductor - superconducting order parameter, its temperature dependence and symmetry, are rather scarce to date.

**2. Phase diagram.** Phase diagram for the alkali-metal pnictides strongly differs from that for the majority of other iron based superconductors. It is widely known that relatively well studied *RE*

---

[1] e-mail: kuzmichevate@lebedev.ru

OFeAs oxypnictides of the 1111 family ($RE$ is a rare-earth metal) as well as the 122 family pnictides $AE$ Fe$_2$ As$_2$ ($AE$ – alkali-earth metal) being in the stoichiometric state, at temperatures about $T_s$: 120 – 150 K undergo a structural transition from tetragonal phase ($T > T_s$) to orthorhombic phase. At lower temperatures $T_m < T_s$ the structural transition is accompanied with antiferromagnetic transition to a spin density wave (SDW) state (for a review, see [5], [6]). At $T_m < T < T_s$, a so called nematic phase emerges in the $RE$-1111 and $AE$-122 family compounds being a nonmagnetic state with broken $C_4$- rotational symmetry in the crystallographic $ab$-planes ($a \neq b$). Superconducting phase showing a "dome" of the critical temperature develops in the tetragonal phase along with SDW and nematicity suppression under pressure or doping. On the contrary, superconductivity emerges in the stoichiometric state in the 111 family alkali-metal pnictides [3, 7–12]. Nonetheless, phase diagram of the 111 family is not universal and changes drastically for the compounds with various alkali metals.

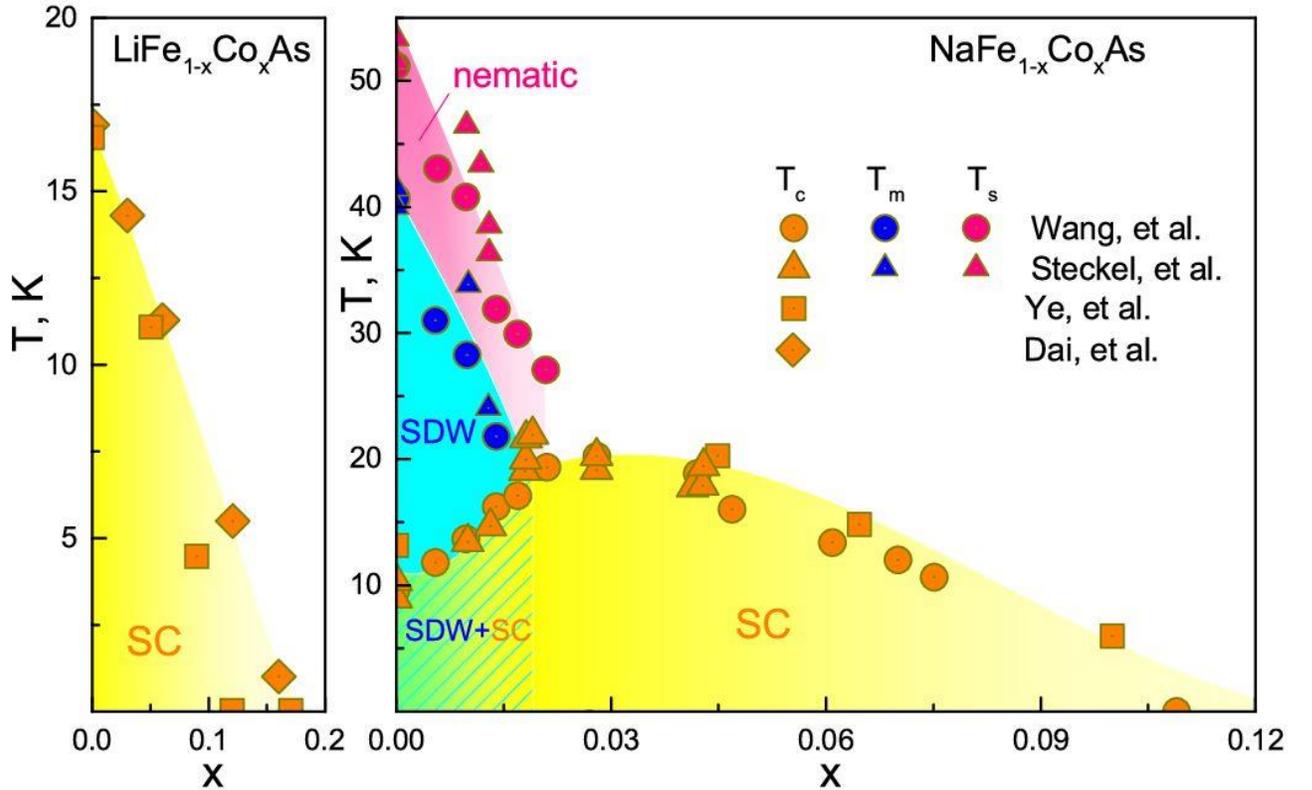

Fig. 1. Doping phase diagrams of LiFe$_{1-x}$Co$_x$As and NaFe$_{1-x}$Co$_x$As. Pink area shows the nematic phase developing at $T_m < T < T_s$, cyan area shows the antiferromagnetic (AFM) phase with a spin density wave (SDW) at $T < T_m$, yellow area depicts the superconducting (SC) phase existing at $T < T_c$. The region of coexistence between AFM and superconducting phases (bulky separated) is shown by dashed green area. The experimental temperature values of the superconducting transition $T_c$ (orange symbols), magnetic $T_m$ (blue symbols), and structural transition $T_s$ (pink symbols) are taken from [3] (triangles), [9] (rhombs), [10] (circles), and [11] (squares)

LiFeAs compound is fully nonmagnetic [13] and naturally has optimal superconducting properties, with maximum critical temperature $T_c = 17 - 18$ K. As shown in [3, 9–11], superconductivity rapidly destroys under partial electron substitution (Fe,Co) within the superconducting FeAs blocks at $x \approx 0.12 - 0.16$ (Fig. 1). Similarly looking phase diagram was obtained for partial substitution by other transition metals (Ni, Cu, V) [14–16], as well as for LiFeAs under pressure, [17] and for the crystals with lithium deficiency [18]. In the latter case, unfortunately, the $T_c$ dependence on the deficiency $\delta$ for $Li_{1-\delta}FeAs$ is not determined reliably to date. Long magnetic order establishes under neither any strong electron, nor any strong hole doping [9]. Noteworthy, such behavior resembles the evolution of magnesium diborides $MgB_2$ properties: superconducting properties of these nonmagnetic layered HTSC are also optimal in the stoichiometric state, whereas under any available substitution (Mg,Al) or (B,C) critical temperature $T_c$ tends to zero [19]. However, a very recent work [20] reported signs of a coexistence of nematicity and superconductivity discussed below.

On the contrary, orthorhombic phase and a magnetic order develop in stoichiometric NaFeAs, although at much lower temperatures, $T_s \approx 55$ K and $T_m \approx 43$ K, respectively [3, 10–12], as compared to the 1111 and 122 family pnictides (see Fig. 1). At $T_c \approx 10$ K, transport, magnetic, and calorimetric probes [3], [10], [11], [21], [22] show a superconducting transition as well. However, by contrast to LiFeAs, a large set of studies [3], [7], [21], [22] shows a natural phase separation in NaFeAs: shunting superconducting clusters related to the tetragonal phase (about 10% of the bulk of the crystal) neighbor with AFM clusters. In (Fe,Co)-doped NaFeAs crystals, the structural and AFM transitions shift toward lower temperatures, with it, the volume of the superconducting fraction increases [12], [21]. Under cobalt doping, maximum $T_c \approx 22$ K is reached in the bulk tetragonal phase as the AFM and nematicity become suppressed, as shown in Fig. 1. Qualitatevily similar phase diagram was obtained in [12] for $NaFe_{1-x}Cu_xAs$ substitution, but there superconductivity vanished much more rapidly with doping (yet for $x \approx 0.05$), whereas the maximum $T_c$ was as low as 12 K.

A presence of nematic fluctuations in LiFeAs and NaFeAs at $T > T_s$ was demonstrated in transport and NMR studies [7], [23].

**3. Band structure and the Fermi surfaces.** For the majority of iron-based HTSC, hole barrels near the $\Gamma$-point and electron barrels are formed near the M-point of the first Brillouin zone, both slightly warped along the $k_z$ direction of the momentum space, and connected by the nesting vector $Q = (\pi, \pi)$ in the 2-Fe unit cell (for a review, see [5, 6, 24, 25]). Band structure calculations for the 111 family pnictides are presented in [26–28]. ARPES-studies [4, 9, 13, 14, 20, 26, 29–36] showed that the Fermi surfaces of the 111 pnictides hardly resembe those for other families, but show quite differences.

In stoichiometric LiFeAs, the radii of the Fermi surface barrels are strongly different (Fig. 2a): at the $\Gamma$-point, a shallow barrel is resolved, whereas the radius of the outer hole barrel is much larger than that of the electron barrels. As a result, nesting at the vector $Q = (\pi, \pi)$ is fully absent in LiFeAs [9, 13–15]. Under electron substitution (Fe, Co), as shown in Fig. 2 taken from ARPES studies [9], the volume of the M-point barrels increases as the volume of the outer hole barrel diminishes. Just for the compound with $x \approx 0.12$ cobalt concentration, the Fermi surface barrels become fully nested (Fig. 2c, h). For hole-doped $LiFe_{1-x}V_xAs$, as reported in [14], with $x$ decrease, only the inner $\Gamma$-point barrel expands (Fig. 3), thus becoming nested with the electron barrels at $x \approx 0.084$ (Fig. 3d).

It is interesting to account that in both cases, as shown in [9], [14], along with nesting in the $\Gamma$-M direction development, superconductivity becomes suppressed: an ideal nesting is reached in strongly overdoped Co-substituted $LiFe_{0.88}Co_{0.12}As$ compound with $T_c \to 0$ (see phase diagram in Fig. 1), and in nonsuperconducting $LiFe_{0.916}V_{0.084}As$ compound (the (Li, V) doping phase diagram is shown in Fig. 1d in [14]). On the other hand, the data obtained in [15] question the universality of such statement: under (Fe, Cu) substitution, no significant Fermi surface reconstruction was observed. Then, such (Fe, Cu) substitution is actually isovalent, which is caused by a localization of the doping electrons, accordingly to the authors' interpretation. Therefore, in order to reveal any correlation between the nesting quality and the $T_c$ value, further studies of the Fermi surface topology in doped

LiFeAs seem of high importance.

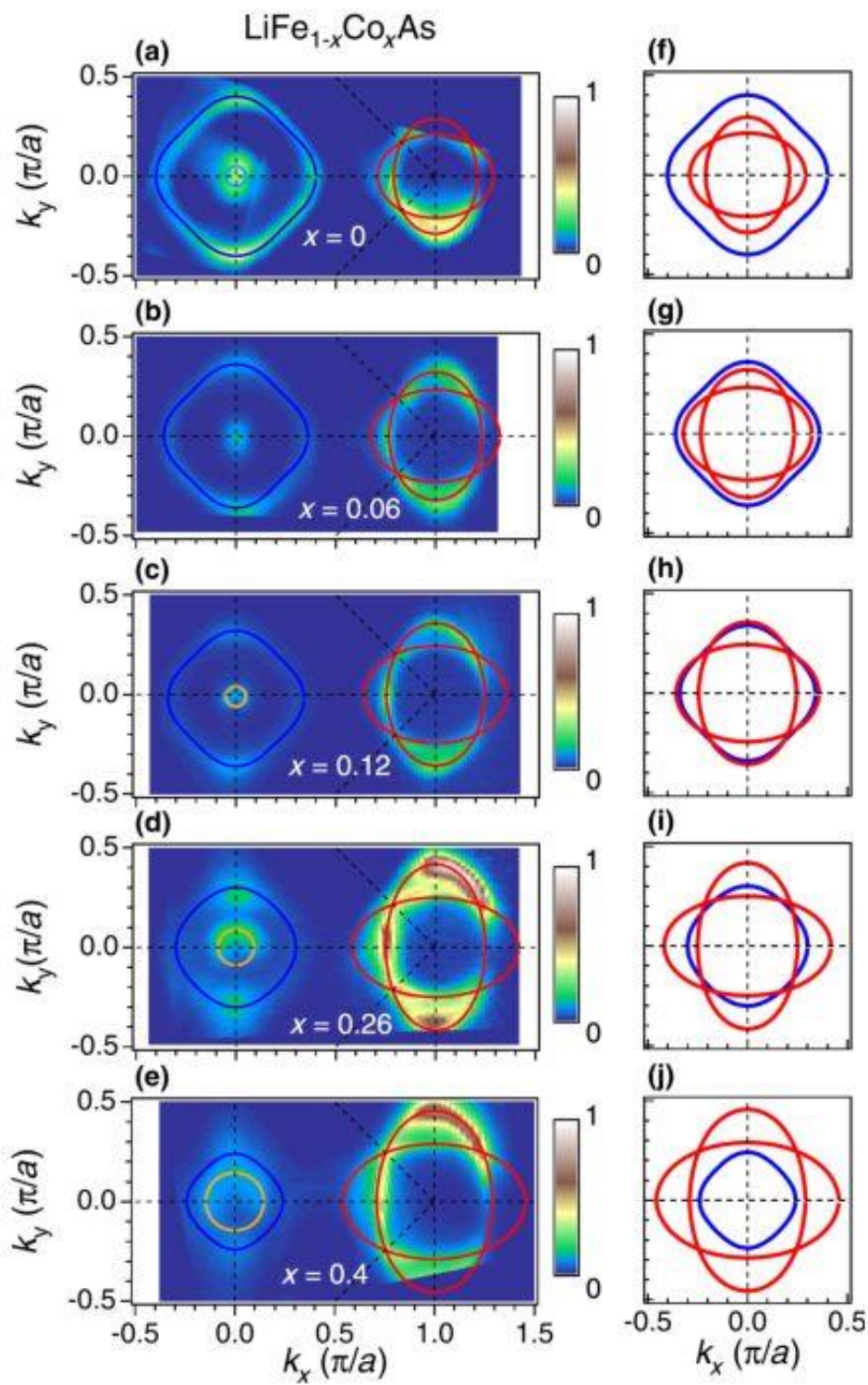

Fig. 2. (a)–(e) – Fermi surface evolution with electron doping for LiFe$_{1-x}$Co$_x$As single crystals with various cobalt concentrationss as determined by ARPES. (f)–(j) – The profiles of the corresponding electron Fermi surfaces (red) as compared to hole ones (blue). The figure is adapted from [9] under CC 3.0 license

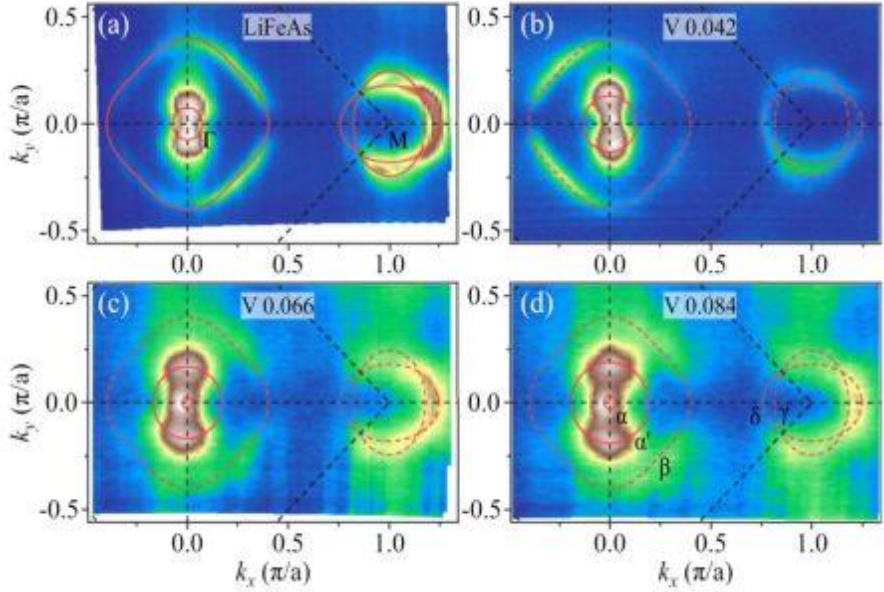

Fig. 3. Fermi surface evolution wit hhole doping for LiFe$_{1-x}$V$_x$As single crystals with various vanadium concentrations as determined by ARPES. Red dashed lines depict the Fermi surface profiles for stoichiometric LiFeAs (the same as shown by solid red lines in (a) panel). Solid lines in (b)–(d) illustrate the profiles of the inner hole barrel for the labelled concentrations $x$. The figure is taken from [14] with the authors' and publisher's permission. © (2021) American Physical Society

Rather interesting results were obtained in recent high-resolution ARPES studies [26,31]. In LiFeAs, as well as in iron-based HTSC of other families, a noticeable (about 10 meV) band splitting in the high symmetry points was detected, caused by a spin-orbit coupling (SOC) [26,31]. In particular, the authors claim that namely SOC is responsible for the shallow Fermi surface barrel emergence near the $\Gamma$-point.

The features of the NaFeAs phase diagram provide an unique opportunity to observe the change in the symmetry of the crystal and band structure in dependence on temperature and doping using ARPES. However, from the experimental point of view, an intermediate problem arises. Generally, a formation of mirror-oriented crystallographic domains (twinning) is typical for single crystals of iron-based superconductors. Since the dimension of the domains is comparable with the ARPES beam diameter, the resulting experimental data is the superposition of the dispersion curves for the both domain orientations, thus making it impossible to resolve an in-plane anisotropy. Several detwinning procedures are reviewed in detail in [37], in particular, a most widely used uniaxial deformation along $a$- and $b$-lattice directions.

Temperature evolution of the stoichiometric NaFeAs Fermi surface is shown in [36,38] (Fig. 4a–h). In the metallic phase with four-fold $C_4$-symmetry, the Fermi surface cross section at $k_z = 0$ represents a circle around the $\Gamma$-point, and two crossing ellipses in the corner of the Brillouin zone (see red profiles in Fig. 4a). The picture remain almost the same under applied uniaxial stress, as shown in (b). There, the nesting condition is satisfied only for several Fermi surface points, along $k_x$ and $k_y$-directions. With temperature decrease, the $C_4$-symmetry breaks: in the nematic phase, a hole barrel distortion along the stress direction, and disappearance of one of the ellipses around the M point are well visible (Fig. 4c–e). Finally, in the AFM phase, the most of the Fermi surface pockets are gapped (Fig. 4g, h).

In superconducting Na(Fe, $Tm$)As, contrary to LiFe(Fe, $Tm$)As, doping does not lead to a full nesting between electron and hole pockets at $k_z = 0$ [32–35], although a possibility of quite ideal nesting for several $k_z$ values was supposed in [32].

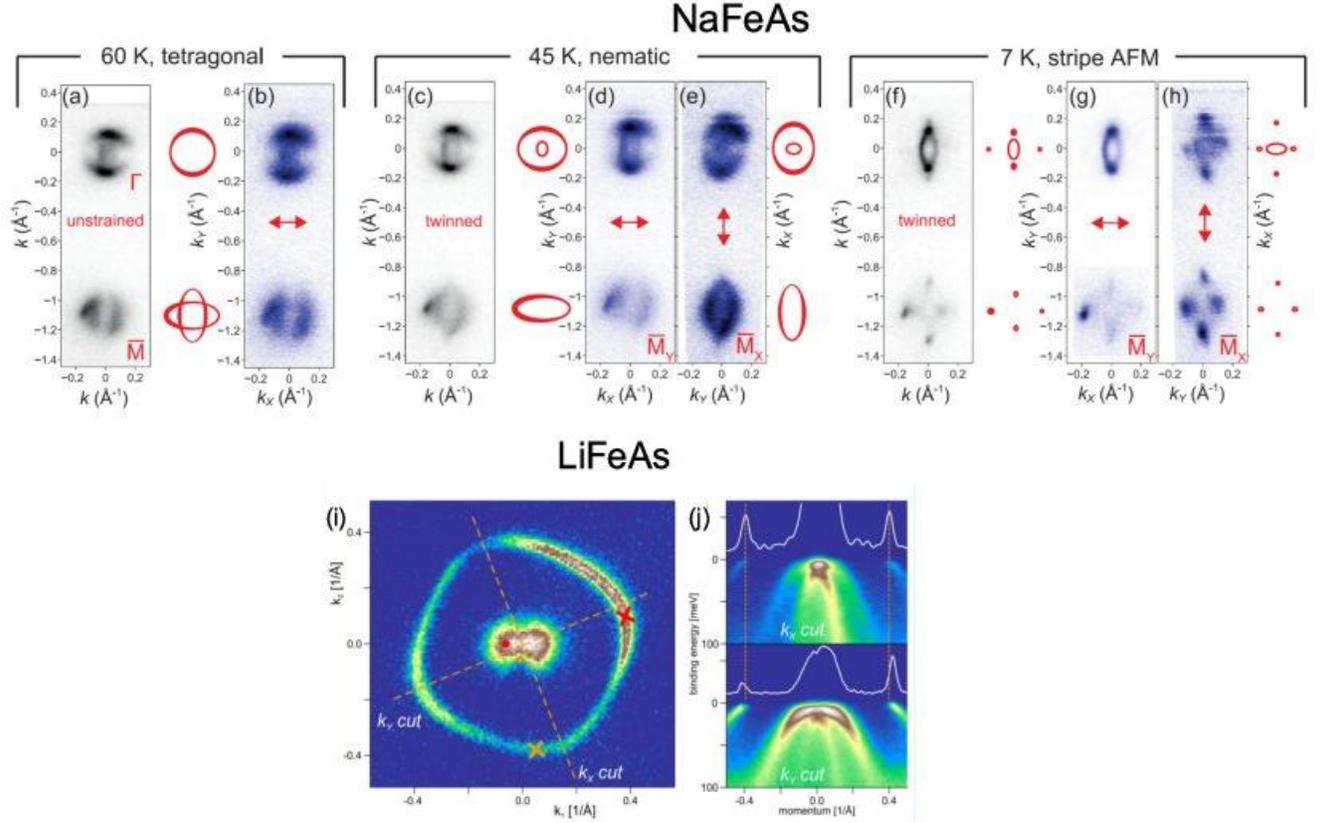

Fig. 4. The Fermi surfaces of stiochiometric NaFeAs near the $\Gamma$ and the M-points as determined using ARPES: (a), (b) – at $T = 60$ K (tetragonal structure, metallic phase); (c)–(e) – at $T = 45$ K (orthorhombic structure, nematic phase); (f)–(h) – at $T = 7$ K (orthorhombic structure, AFM phase). The data for twinned crystals are shown in x (a), (c), (f) panels; similar data for uniaxially strained samples (red arrows indicate the strain direction) are presented in (b), (d), (e), (g), (h). Red lines show the extracted profiles of the corresponding Fermi surfaces. Note the $k_x$- and $k_y$- axes are swapped in (d), (e) and (g), (h) panels. (i) – Hole barrels around the $\Gamma$-point in twinned LiFeAs single crystal in the superconducting phase at $T = 7$ K. The band structure cuts along $k_x$ and $k_y$ directions (dashed lines in (i)), as well as the corresponding energy dispersion curves (solid lines) are shown in (j) panel. Vertical dotted lines expose the Fermi surface anisotropy in the $xy$-plane of the momentum space. Figures (a)–(h) are taken from [36] (under the CC 4.0 license), panles (i), (j) are reprinted from [20] with the authors' and publisher's permission. © (2021) American Physical Society

A minor distortion (about 4 %) of the outer hole barrel at the center of the Brillouin zone was resolved in very recent ARPES probe [20] of twinned stoichiometric LiFeAs below $T_c$ (Fig. 4i, j). The reasons facilitationg the observation of the $ab$-plane anisotropy without applied uniaxial stress at all are discussed by the authors. Anyway, such $C_4$-symmetry breaking below $T_c$ could indicate a superconductivity development in the nematic phase, that was not observed earlier in any high-temperature superconductor excepting FeSe. Moreover, since the asymmetry of the band structure resolved at low tempertures vanishes above $T_c$, the authors of [20] suppose superconductivity responsible for the observed nematicity in LiFeAs.

**4. On the theoretical models describing the superconducting state, and the attempts of their experimental verification in the 111 family.** One of the earlier theoretical works [39] shows that despite observation of partial isotope effect [40], electron-phonon coupling appears rather weak in the iron-based superconductors, thus being insufficient to provide their relatively high $T_c$. Later, in order to describe the superconducting mechanism in the iron pnictides and chalcogenides, several theoretical models were suggested. In the framework of spin-fluctuation models, where the Cooper pairs are formed via the nesting between the Fermi surface parts of the same kind of orbitals (so called "intraorbital" pairing), it is possible to obtain a sign-reversal superconducting order parameter of $s^\pm$-type [41–45] (formally, negative superconducting gap $\Delta < 0$ for one Cooper pair condensate implies the phase of its wave function is shifted by $\pi$ as compared to that of another condensate), or complex $s + is$-type [46] (with arbitrary phase shift between superconducting condensates, different from $\pi$, and broken time reversal symmetry). Spin resonance at the nesting vector $Q = (\pi, \pi)$ was widely observed in the superconducting state in inelastic neutron scattering experiments with iron-based superconductors of various families (for a review, see [44,47,48]). Also it is worth noting that some tunneling probes using break-junction technique [49–51] and point contact Andreev reflection (PCAR) technique [52,53], a resonant interaction between Andreev current and a characteristic bosonic mode, possibly a spin exciton, was observed in the superconducting state: the boson energy at $T = T_c$ did not exceed an indirect superconducting gap $\Delta_L(0) + \Delta_S(0)$ or $2\Delta_L(0)$, thus satisfying the spin resonance condition in accordance with the calculations [54,44].

On the other hand, coupling through nematic fluctuations [56] or orbital fluctuations enhanced by phonons [57-59], a strong intraband electron-phonon interaction [60], as well as accounting for a Fano-Feshbach resonance near a Lifshitz transition or a BCS-BEC crossover [61] lead to a so called $s^{++}$-symmetry of the superconducting order parameter without sign change (i.e. the wave functions of all the superconducting condensates are in phase). At once, spin fluctuations could be considered as an additional pairing channel responsible for a strong anisotropy of the superconducting order parameter, even sign-reversal[56,58,60].

Due to the Fermi surface features of LiFeAs [9, 13–15], in general, a pronounced spin resonance at the $(\pi, \pi)$ is hardly expected. Indeed, in the only work [62] a weak spin-resonance maximum was detected below $T_c$ at the energies $\varepsilon_0 \approx 6 - 11$ meV and at the $Q$ vector. The authors of [63] have also obderved some resonance with energy $\varepsilon_0 \approx 5$ meV, however, do not speculate on its origin. On the other hand, theoretical calculations [45,58,59] predict a robust gap solution with $s^\pm$-symmetry even in case of a poor nesting in the $\Gamma$-M-direction, whereas the gap with the smallest absolute value supposed to develop at the smallest Fermi surface pocket. Accounting orbital selectivity (different correlation strengths in the bands formed by different orbitals; in particular, Cooper pairing strength) within the $s^\pm$-approach, the authors of [64] qualitatively reproduced the gap structures obtained in [45,58,59] for "pure" $s^\pm$-case, excepting that the condensate with the large gap developed at the inner hole barrel below $T_c$.

An existence of the bosonic mode in LiFeAs is under discussion now. Optical studies [65] have not reported any overgap feauters those would be typical for the superconducting state below $T_c$. In the incoherent multple Andreev reflections effect (IMARE) spectroscopy studies of LiFeAs single crystals, a fine structure caused by a resonant coupling with a bosonic mode was reproducibly absent [66,67], although it was well-resolved in the 1111 oxypnictides of various composition [49–51]. In the scanning tunneling microscopy (STM) probes [68–70], an overgap dip-hump structure appeared in the obtained $dI(V)/dV$-spectra was interpreted withinin the approach [68,70] as a footprint of electron density of states (DOS) renormalization caused by a spin resonance. On the other hand, in the works [71–73] a set of arguments against such interpretation was suggested. In particular, the authors of [72] attribute the dip-hump structure to a coupling with nematic fluctuations, whereas the authors of [73] show that similar dip-hump structures usually appear in the $dI(V)/dV$-spectra of tunneling junctions owing to surface defects influence.

Theoretical calculations [58,59] showed that a moderate anisotropy (including that along the $k_z$-direction) of the superconducting gaps developing below $T_c$ at electron and hole barrels could be obtained even accounting $s^{++}$-interaction between the Fermi surface parts formed by different orbitals solely (so called "interorbital" coupling). Moreover, when combining the pairing channels via spin and orbital fluctuations, it is possible to simulate almost any kind of gap anisotropy, even nodal (turning to zero at certain momenta $\Delta(k) = 0$) or sign-reversal. The most important results of these studies [58,59] are: (a) the largest superconducting gap developing at the smallest Fermi surface pocket (the inner barrel around the $\Gamma$-point) at the strong $s^{++}$-interaction limit; (b) if the strengths of the $s^{++}$ and $s^{\pm}$-interactions are comparable, the "negative" superconducting order parameter develops only at the outer hole barrel, whereas $|\Delta| \mapsto 0$ for this band.

Multiple-gap superconducting state of NaFeAs could be described in the both frameworks, a univeral $s^{\pm}$-approach [74], and accounting orbital selectivity [75]. An observation of a weak spin resonance with energy $\varepsilon_0 / k_B T_c = 4 - 6$ in underdoped and optimally doped Na(Fe,$Tm$)As ($Tm$ = Co, Cu) below $T_c$ was reported in [12, 76–79]. In overdoped NaFe$_{0.92}$Co$_{0.08}$As, according to [78], despite the significantly lower $T_c$ (as compared to that for the optimal composition), the $\varepsilon_0$ energy remained almost unchanged. Interestingly, for the compounds close to the stoichiometric NaFeAs composition, inelastic neutron scattering experiments [77,78] resolved a double spin resonance at the same nesting vector, whereas the second (low-energy) maximum at $\varepsilon_0^{min} \approx 3$ meV (corresponding to a spin gap opening) vanished along with the AFM order suppression under temperature or dopant concentration increase.

Recent theoretical calculations [80] revealed a presence of an ordered state originated from interorbital coupling at the edge of the nematic phase in the superconductors having a hole pocket formed by $d_{xy}$-orbitals (in particulr, NaFeAs and Ba-122). The predicted phase could strongly affect the superconducting properties, as well as cause electron DOS humps in the vicintiy of the Fermi level (resembling the pseudogap humps widely discussed in HTSC cuprates [81]). The pseudogap-like nonlinearity of the electron DOS was observed in ARPES [82] and tunneling experiments [83] with the Ba-122 family superconductors. For NaFeAs, similar data are not published yet.

**5. The structure of the superconducting order parameter studies.** The detailed studies of the superconducting order parameter (the dependence of the Cooper pair coupling energy $2\Delta_i$ on the magnitude and the direction of the Fermi momenta) in lithium-based pnictides are not numerous. The available data are obtained for nominally stoichiometric composition LiFeAs crystals, whereas doped compounds Li(Fe,$Tm$)As are not studied at all. In the majority of the works, the authors report a multiple-gap superconductivity in LiFeAs (a coexistence of several condensates with different Cooper pair coupling energies) with nodeless superconducting gaps. Noteworitly, since phase-sensitive probes (such as, for example, Josephson spectroscopy) have not been made for the 111 family pnictides to date, the available data cannot distinguish directly a change of the sign of the superconducting order parameter (i.e. whether the phase shift between the condensates' wave functions takes place).

Figure 5 shows the dependence of the characteristic ratios $2|\Delta_i(0)|/k_B T_c \equiv r_i^{BCS}$ on the critical temperature basing on the data obtained in literature. The critical temperatures $T_c < 18$ K obtained in the most of papers for the single crystals of nominal LiFeAs composition seems very likely caused by a local lithium deficiency, hence, hereafter we will use the Li$_{1-\delta}$FeAs notation.

ARPES-experiments [20], [29,30] revealed a presence of at least three superconducting condensates. Despite the absolute values $\Delta_i(0)$ are little contradictory (circles in Fig. 5, right panel), the data are qualitatively similar: at $T < T_c$ the largest superconducting gap develops at the inner hole barerl, whereas the smallest gap develops at the outer one. Additionally, the order parameters have a valuable anisotropy in the momentum space. Such gap structure agrees well with the calculations [58,59] within the $s^{++}$-approach (without sign change), as well as could be reproduced in the framework of $s^{\pm}$-model accounting orbital selectivity [64] (for a sigh-reversal order parameter).

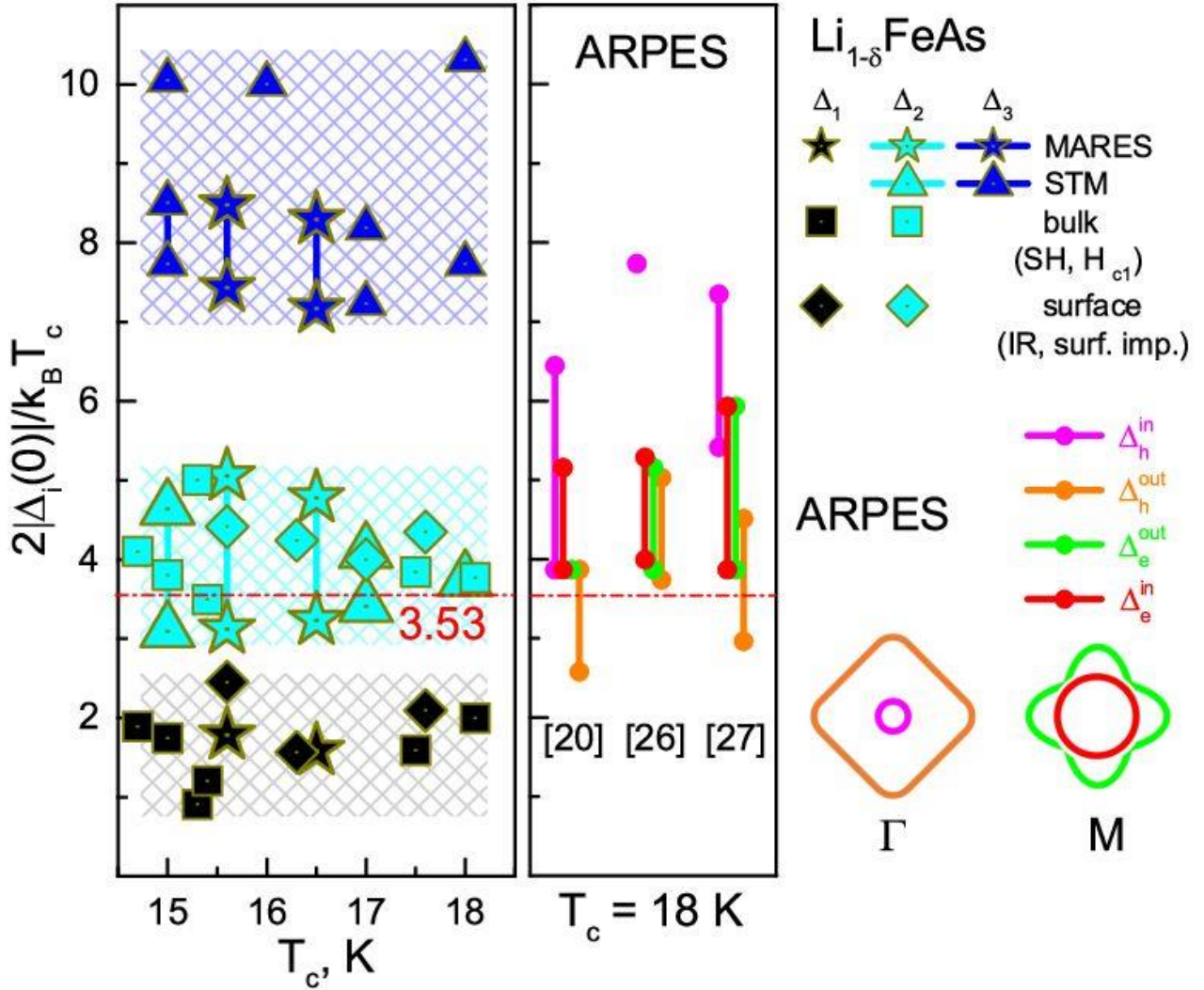

Fig. 5. The dependence of the characteristic ratios of the supernducting order parameters $2\Delta_i(0)/k_BT_c$ on the critical temperature $T_c$ for $Li_{1-\delta}$FeAs using the data measured by incoherent multiple Andreev reflection effect (IMARE) spectroscopy (stars) [67],[2], scanning tunneling microscopy (STM, triangles) [68–72, 84, 85], bulk probes (specific heat and lower critical field measurements, squares) [86–90], and surface probes (infrared (IR) reflection specrtoscopy, surface impedance measurements, rhombs) [65,91]. Connected symbols illustrate the anisotropy degree of the superconducitng order parameter. Dashed areas cover the ranges of the values obtained. ARPES data (circles) [20], [29,30] obtained for LiFeAs with one and the same $T_c \approx 18$ K, are shown on the right and shifted horisontally for clarity. The ARPES data points are colored similarly to the corresponding electron and hole Fermi surface barrels sketched. Dash-dot line shows the weak-coupling BSC limit

In the electron bands [20], [30] superconducting order parameters have midlle, almost similar magnitudes. Although in earlier studies [29,30] four-fold ($90°$ rotation symmetry) angle dependences of the gaps $\Delta_i(\theta)$ were supposed (where $\theta$ is the angle in the $k_x k_y$-plane), wery recent high-resolution probes [20] showed that $\Delta(\theta)$ could be well fitted by a two-fold function ($180°$ rotation symmetry). The latter corroborates the authors' claim [20] about the superconductivity development in the nematic phase.

The data obtained by tunneling [67–72, 84, 85], bulk [86–90], and surface probes [65,91] summarized in Fig. 5 are obviously grouped into three bunches with the characteristic ratio ranges ($0.9 - 2.4$), ($3.1 - 5.1$), and ($7.2 - 10.3$). Hereafter, the supercnducting order parameters are denoted as $\Delta_1$, $\Delta_2$, and $\Delta_3$, respectively. Using IMARE spectroscopy [67], three distinct superconducting order parameters were directly measured (stars in Fig. 5): the small gap with the characteristic ratio $r_1^{BCS} \approx 1.7 = 3.53$ showing no signs of anisotropy, rather strongly anisotropic middle gap with $r_2^{BCS} \approx 3 - 5$, and minor splitted large gap with $r_3^{BCS} \approx 7.3 - 8.4$ (the value ranges correspond to the anisotropy degree, about 40 and 14%, respectively). Aside from the above mentioned data, an anisotropy of the superconducting order parameters was resolved only in [84] by quasiparticle interference technique (connected triangles in Fig. 5): clearly, this data agrees well with the IMARE results [67] in both, the anisotropy degrees of the large and the middle gap, and in the values of their characteristic ratios.

We note that in STM experiments [68–72, 84, 85], only two superconducting gaps are observed, the large and the middle ones (triangles in Fig. 5). For the large gap, the values $2\Delta_3(0)/k_B T_c \approx 7.2 - 8.5$ estimated in [68,69,85] correspond to the $\Delta_3$ anisotropy range determined in [67,84]. The characteristic ratio of the order parameter developing at the inner hole barrel estimated in [30] lays within this range as well, whereas similar data from [20,29] appear a bit lower. At the same time, another STM probes [70–72] report higher characteristic ratio values, up to 10.3. One could suppose, since the superconducting gap magnitudes in [70–72] were estimated directly from the positions of the tunneling maxima in the $dI(V)/dV$-spectra, an influence of inelastic processes characterized by the broadening parameter $\Gamma = \hbar/2\tau$ (where $\tau$ is the typical energy relaxation time) is a reason of such gap overestimation. A finite $\Gamma$ value generally leads to both, the broadening of the DOS features and the gap-edge DOS peaks shifting toward higher energies. According to recent ARPES studies [92], in LiFeAs the $\Gamma$ value could be extremely high, even comparable with $\Delta(0)$.

Temperature dependences of the superconducting gaps in Li$_{1-\delta}$FeAs are obtained using only two experimental methods to date. The dependences of the three gaps $\Delta_{1,2,3}(T)$ directly determined using IMARE spectroscopy are typical for the case of a moderate interband coupling, whereas the anisotropy degrees of the $\Delta_{2,3}$ remain almost constant within a wide temperature range [67]. Similarly looking temperature behavior of the large and the middle gaps $\Delta_{2,3}(T)$ was obtained in [69] using fitting with the Dynes model of the $dI(V)/dV$-spectra measured at $T < T_c$.

On the other hand, bulk studies (measurements of the specigic heat and the lower critical field) [86–89] as well as surface studies (infrared (IR) reflection spectroscopy, surface impedance measurements) [65,91] report an observation of the middle and the small gap (squares and rhombs in Fig. 5). Possibly, it relates with a low Cooper pair concentration in the condensate with the large gap, as shown in [67]. Nonetheless, the observed data diversity $r_2^{BCS} \approx 3.5 - 5.0$ for the middle gap (cyan symbols in Fig. 5) matches exactly its anisotropy range determined in [67,84], and also agrees with the characteristic ratio ranges for the anisotropic order parameters in the electron band estimated using ARPES [29,30]. As well, one should not exclude a possible anisotropy of the small gap (accounting that in ARPES probes, pairing anisotropy was observed in all the bands, see connected circles in Fig. 5, right panel) as a reason of the scattering of its characteristic ratio $1.0 - 2.5$ (black symbols in Fig.

5). However, ARPES estimates provide a bit higher values $r_1^{BCS} > 2.6$.

Extremely few experimental studies of the sodium-based 111 family pnictides gap structure have been made at the moment. The available data are obtained for cobalt-doped Na(Fe, Co)As barely and summarized in Fig. 6. The experimental points (blue and black) are grouped into two bunches with $2\Delta_i(0)/k_BT_c \approx (2.6 - 4.5)$ and $(5.3 - 8.8)$, both shaded in Fig. 6. Although it may seem that the characteristic ratios show a tendency to decrease toward the overdoped region, the available amount of data is of course insufficient to prove it.

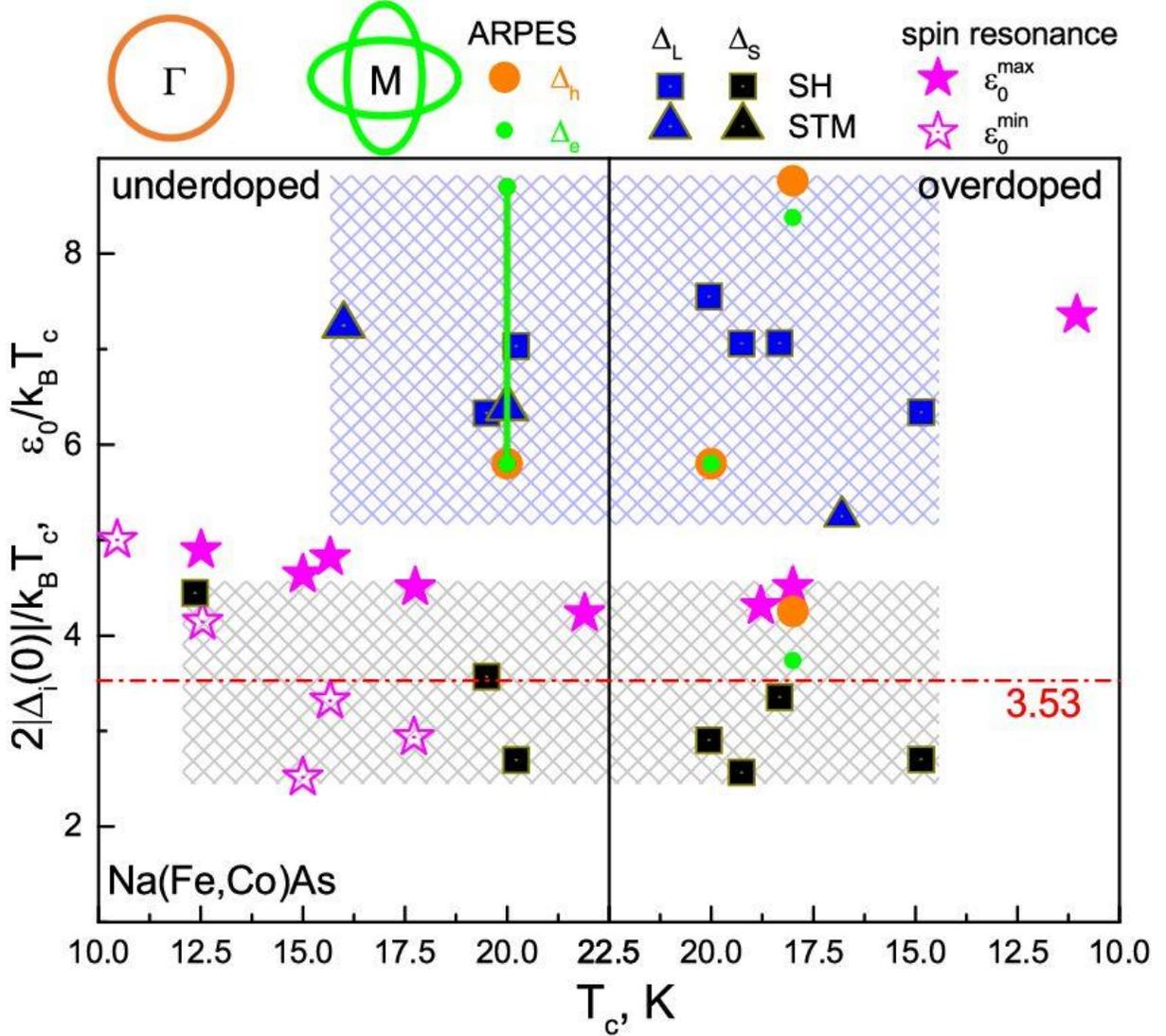

Fig. 6. The dependence of the characteristic ratios of the superconducting gaps $2\Delta_i(0)/k_BT_c$ determined using STM (triangles) [94,95], specific heat (squares) [12], [10], and ARPES (circles) [32, 33, 93] versus critical temperature $T_c$ in doped Na(Fe, Co)As. The circles are colored in accordance with the corresponding electron and hole Fermi surface pockets sketched. The connected symbols illustrate the anisotropy degree of the superconducting order parameter. Dashed areas cover the ranges of the obtained values of the superconductign gap characteristic ratios. For comparison, the dependence of the characteristic ratio of the spin resonance energy $\varepsilon_0/k_BT_c$ (solid stars for the high-energy mode, open stars for the low-energy one) versus $T_c$, as determined in inelastic neutron scattering probes [76–78]. Dash-dot line borders the weak-coupling BCS limit

Specific heat measurements of Na(Fe, Co)As samples with various doping degree [10], [12] (squares in Fig. 6) showed a presence of the large superconducting gap with $2\Delta_L(0)/k_BT_c \approx 6.3 - 7.5$ well-exceeding the weak-coupling BCS limit, as well as the small gap with the characteristic ratio $2.6 - 3.6$. With it, similar data [12] obtained for underdoped sample from the same batch with $T_c \approx 12.3$ K were well fitted with a single-gap model.

The authors of the ARPES studies of Na(Fe, Co)As [32,33,93] agree that a uniform superconducting condensate develops at two electron barrels of the Fermi surface. Nonetheless, the exact values of the energy parameters are rather contradictory (circles in Fig. 6). The two studies of overdoped crystals [33,93] reported that the superconducting gaps developed in the electron and the hole bands are quite similar, but the corresponding characteristic ratios determined in [33,93] are rather different (green and orange symbols in Fig. 6, right panel): the lowest value $r_S^{BCS} \approx 4$ [32] agrees well with that of the small gap estimated using a specific heat temperature dependence [12] ,[45], whereas the highest values $r_L^{BCS} \approx (5.8 - 8.8)$ [33,93] lays within the range for the large gap determined using STM [10,12,94,95]. For an underdoped composition (left panel of Fig. 6), a strong anisotropy of the large order parameter developing at the electron pockets was resolved using ARPES [93], but was not been observed by that group for an overdoped crystal with similar $T_c$. An appearance of the superconducting order parameter anisotropy within the region where AFM and superconducting phases naturally coexist, may indicate a possible SDW influence to the gap structure, thus needing in further detailed studies with high energy resolution.

Among the other garoups, the superconducting gap anisotropy is discussed now. The specific heat data $C_{el}(T)$ [10], [12] and thermal conductivity measurements [96] are well fitted with nodeless isotropic order parameters. On the other hand, the authors of [12] do not exclude an anisotropy of the superconducting properties, if accounting the limitations of the model used to fit the $C_{el}(T)$ data with. At the same time, the temperature dependence of the Cooper pair concentration (determined using the London penetration depth measurements) is claimed [97] to be fitted in the framework of anisotropic superconducting gaps only for both, underdoped and overdoped compositions. Contrary, the tunneling spectra obtained in [95] could be fitted with the Dynes model for both cases, isotropic and anisotropic gap, by varying the broadening parameter $\Gamma$. Unfortunately, the tunneling data for Na-111 are almost absent: the available probes made by STM [94,95] have resolved only a single superconducting gap, whereas Andreev spectroscopy experiments have not been made at all.

In order to compare with the gap parameters, the characteristic ratios of the spin resonance energy $\varepsilon_0/k_BT_c$ are also shown in Fig. 6 (stars), those determined in inelastic neutron scattering studies of Na(Fe, Co)As [76–78]. As mentioned above, the double spin resonance is reproducibly observed in underdoped region [76–78] (left panel of Fig. 6): the energy of the first resonance mode $\varepsilon_0^{min}$ (low-energy mode existing in the AFM phase at $T < T_m$) keeps almost independent with $T_c$, thus resulting in the increase of its characteristic ratio when approaching the AFM phase (open stars in Fig. 6). For the second resonance (observed in the superconducting state only), the value $\varepsilon_0^{max}/k_BT_c \approx 4.2 - 4.9$ weakly depends on doping (solid stars in Fig. 6), excepting a single droped off point on the right. In Na(Fe, Cu)As crystals with slightly overdoped composition and $T_c \approx 12$ K, a bit higher ratio $\varepsilon_0/k_BT_c \approx 5.5$ was obtained in [12]. Noteworthily, the $\varepsilon_0^{max}/k_BT_c$ data shown in Fig. 6 are located almost in the middle between the characteristic ratios for the large and the small gaps; more precisely, $\varepsilon_0^{max}$ is about $\approx 2.5\Delta_S(0)$ being less than $\Delta_L(0) + \Delta_S(0)$. Therefore, in accordance with the available data statistics shown in Fig. 6, the spoin resonance condition predicted theoreticall in [54,55] is satisfied in Na(Fe, Co)As.

**6. Conclusion.** The above brief review shows that the studies of the alkali-metal-based 111

family superconductors, discovered about 13 years ago, are far from being finished yet. The available experimental data are rather contradictory and mostly inconsequtive. Nonetheless, due to their extraordinary properties those are not typical for other families of iron-based superconductors, namely the $A$-111 family pnictides could be crucial for answering the majority of fundamental questions. The most experimental problems seem as follows:

- A direct measurement of the structure of the superconducting order parameter (the number, magnitudes, symmetries, and the characteristic ratios of the superconducting gaps, their temperature dependences, and a possible phase shift $s + is$) in doped $A\text{Fe}_{1-x}Tm_x\text{As}$ ($A$ = Li, Na; $Tm$ = Co, Ni, Cu, V, Rh), as well as in the compounds with alkali metal deficiency $A_{1-\delta}\text{FeAs}$, with various $x$ and $\delta$. A comparison between the properties of underdoped and overdoped crystals with electron, hole, and isovalent substitution, and revealing the evolution of their properties along the corresponding doping phase diagrams.
- Inelastic neutron scattering studies of the spin resonance (the determination of the energy $\varepsilon_0$, its characteristic ratio, and temperature dependence) in overdoped $\text{NaFe}_{1-x}\text{Co}_x\text{As}$, as well as in Na-111 with substitution for other transition metals (Cu, Rh и др.) or with Na deficiency within the whole substitution range available. The uncovering of the nature of the bosonic mode, its energy, and temperature dependence using high-resolution tunneling probes.
- A further detailed study of a possible coexistence between nematicity and superconductivity in LiFeAs, as well as in other iron-based superconductors.

An experimental verification of the above mentioned issues seems to define, to what extent the features of the band structure, magnetism and nematicity influence the superconducting subsystem. Without any doubts, the solving of the listed problems would facilitate an adaptation and generalization of the theoretical models in order to describe the physics of iron pnictides and chalcogenides. Hopefully, this will bring the researchers to the answer of the central question: whether the mechanism of unconventional superconductivity is universal for different families of pnictides and chalcogenides?

The authors are grateful to I.V. Morozov for the high-quality LiFeAs samples provided, moral and technical support, and permanent regard, as well as M.M. Korshunov and V.M. Pudalov for fruitful discussions. The work is supported by the state assignment of the Ministry of Science and Higher Education of the Russian Federation (topic "Physics of hightemperature superconductors and novel quantum materials," Grant No. 0023-2019-0005). The experimental IMARE study of LiFeAs with the participation by T.E. Kuzmicheva was done in the framework of Russian Science foundation project no. 19-72-00196.